\documentclass{iopart}
\pdfoutput=1
\usepackage{amssymb}
\usepackage{graphicx}
\usepackage[colorlinks,
citecolor=blue,linkcolor=red,urlcolor=blue]{hyperref}

\def\be{\begin{equation}}
\def\ee{\end{equation}}

\def\bea{\begin{eqnarray}}
\def\eea{\end{eqnarray}}
\def\non{\nonumber \\}
\def\o{\over}

\def\k{\kappa}

\begin{document}

\begin{flushright}
YITP-16-57
\end{flushright}

\title[One-body reduced density matrix of trapped impenetrable anyons]{One-body reduced density matrix of trapped impenetrable anyons in one dimension}

\author{Giacomo Marmorini$^1$, Michele Pepe$^2$, Pasquale Calabrese$^3$}
\address{$^1$ Yukawa Institute for Theoretical Physics, Kyoto University, Kyoto 606-8502, Japan
and Research and Education Center for Natural Sciences, Keio University, Kanagawa 223-8521, Japan} 
\address{$^2$ Dipartimento di Fisica dell'Universit\`a di Pisa, 
56127 Pisa, Italy} 
\address{$^3$ SISSA and INFN, via Bonomea 265, 34136 Trieste, Italy} 
\ead{giacomo@yukawa.kyoto-u.ac.jp}

\begin{abstract}
We study the one-body reduced density matrix of a system of $N$ one-dimensional impenetrable anyons trapped 
by a harmonic potential. 
To this purpose we extend two methods developed to tackle related problems, namely the 
 determinant approach and the replica method. 
While the former is the basis for exact numerical computations at finite $N$, the latter has the advantage of 
providing an analytic asymptotic expansion for large $N$. 
We show that the first few terms of such expansion are sufficient to reproduce the numerical results  to an excellent accuracy 
even for relatively small $N$, thus demonstrating the effectiveness of the replica method.
\end{abstract}


\maketitle

\section{Introduction}
Quantum statistics is among the most fundamental concepts in physics and is at the basis of the existence of formidable physical systems at all scales, from Bose-Einstein condensates  to neutron stars. 
While in three (and higher) spatial dimensions, bosonic and fermionic statistics exhaust all the possibilities, physicists 
have postulated and then investigated the existence of (quasi-)particles with generalized statistics, or anyons, in lower dimension for some decades \cite{anyons}. 
In two dimensions, the fractional statistics of the elementary excitation of the quantum Hall effect has certainly represented a huge physical motivation in this direction \cite{fqhe}. In one dimension, numerous  models of anyons have been proposed 
(\cite{kundu-99,amico-99,liguori-99, zhu-96,batchelor-06,patu-07,bf-08,hao-08,mgk-08,bcm-09,patu-09,spk-12,yao-12,patu-15,tep-15,zinner-15} 
among others), 
but most of them have been thought to be essentially playgrounds for  theoreticians for quite a long time. 
This perspective seems bound to change in view of the new experimental achievements, in particular concerning cold atoms in optical lattices. In fact, in the cold atom framework a number of proposals have been formulated that indicate the realization of one-dimensional anyons as feasible within the current experimental techniques. Among those proposals,  the one elaborated by Keilmann et al. \cite{keilmann-11} and subsequently refined by Greschner and Santos \cite{greschner-15}  appears to be particularly promising. Their idea starts from ordinary bosonic or fermionic atoms and  employs a Raman-assisted tunneling process to induce an effective occupation-dependent hopping; the system obtained in this way can be shown to be equivalent to the anyonic Hubbard model, in which the statistical parameter and the on-site interaction can be changed continuously.  
More recently, another very interesting experimental scheme has been suggested by Str\"ater et al. \cite{strater-16}, which makes use of the so-called lattice shaking technique combined with a static tilt of the potential in order to realize the occupation-dependent tunneling. This scheme does not require any additional lasers other than the ones that form the lattice nor any assumptions on the internal atomic structure. 

This kind of developments give reasonable expectations that the knowledge of generalized statistics and its implication on quantum many-body physics  collected in extensive studies of one-dimensional models can be progressively tested against experiments.  In fact, one-dimensionality brings about a wealth of theoretical tools that are unavailable in the study of quantum many-body physics in higher dimension, the Bethe Ansatz being a prominent example. However, despite the fact that the Bethe Ansatz solution provides information about the spectrum, the thermodynamic properties, etc., in general it  does not make the calculation of correlation functions straightforward, so that one has to ultimately rely on extensive numerical computation \cite{ccs-06}. Among the most interesting quantities is the one-body reduced density matrix (or off -diagonal correlation), that is defined as
$$
\rho_N(t,t') = N \int d^{N-1}x \, \bar{\psi}_N(t,x_2,\ldots,x_N) \psi_N(t',x_2,\ldots,x_N),
$$
where $\psi_N$ is the $N$-body ground-state  wavefunction,  and gives access to fundamental physical observables such as momentum distribution, one-particle entanglement entropy, natural orbitals and their occupation, etc. 
Whereas its analytic determination remains a challenging task in general (see  \cite{ccs-06,kk-12}), 
there is at least a class of models, namely models of impenetrable particles (or Tonks-Girardeau models) in various geometries 
(circle, interval with Dirichlet or Neumann boundary conditions, harmonic well), in which this problem has enjoyed a sensible progress over the last years. 
This model is a limiting case of the Bethe Ansatz integrable Lieb-Liniger model of bosons \cite{ll-63} (or anyons) in which the  zero-range mutual interaction ($\delta$-interaction) is infinite (impenetrable, or hard-core, limit). 
The relative simplicity of the Tonks-Girardeau model lies mainly in the fact that the $N$-body ground-state  wavefunction can be related to the one of free spinless fermions via the boson-fermion \cite{girardeau-60} (or anyon-fermion \cite{girardeau-06}) mapping. 
Starting from this, it has been shown that $\rho_N(t,t')$ in the bosonic case can be exactly expressed as the determinant of a matrix whose symmetry depends on the geometry, {\it e.g.} Toeplitz type for circular geometry, Hankel type for harmonic well, etc. \cite{forrester-02a}. 
The generalization of this construction to anyons on a circle has been performed in Refs. \cite{santachiara-07,santachiara-08,patu-07}. Besides, in a remarkable work \cite{gangardt-04}, Gangardt was able to find the full large distance expansion of $\rho_N(t,t')$ in a closed form for bosons on a circle and in a harmonic well; this expansion, which is valid for distances larger than $r_0/N$, where $r_0$ is a characteristic length depending on the geometry, has the advantage of being analytic and, in practice, 
yielding very accurate results  also for finite $N$. 
Building upon this,  a prescription to apply the replica method to anyons in circular geometry has been proposed 
and tested in Ref. \cite{calabrese-08}.

The purpose of this paper is to extend the  study of the one-body reduced density matrix to Tonks-Girardeau anyons trapped by a harmonic potential within the framework described above, possibly providing a theoretical tool for future cold atom experiments which 
are indeed mainly performed in the presence of the harmonic trapping potential.
 First we derive the exact expression of the one-body reduced density matrix in terms of a Hankel determinant in Sec.~\ref{sec:hankel}. Afterwards, in Sec.~\ref{sec:replica}, we determine its complete asymptotic expansion by combining the replica method of \cite{gangardt-04} and the anyonic prescription of \cite{calabrese-08}. Section~\ref{sec:numerics} is devoted to the presentation of the numerical results and the comparison of the two approaches for some choices of the parameters. Concluding remarks are given in Sec.~\ref{sec:concl}.

\section{The one-body reduced density matrix as a Hankel determinant} \label{sec:hankel}

Anyonic statistics in one dimension can be defined by introducing field operators with the following commutation relations:
\bea
\Psi_A^\dag(x_1)\Psi_A^\dag(x_2)&=&e^{i\k \pi\epsilon(x_1-x_2)} 
\Psi_A^\dag(x_2)\Psi_A^\dag(x_1), \non
\Psi_A(x_1)\Psi_A^\dag(x_2)&=&e^{-i\k \pi\epsilon(x_1-x_2)} 
\Psi_A^\dag(x_2)\Psi_A(x_1) +\delta(x_1-x_2),
\label{ancomm}
\eea
where $\epsilon(z)=-\epsilon(-z)=1$ for $z>0$ and $\epsilon(0)=0$. $\k$ is called statistical
parameter and equals $0$ for bosons and $1$ for fermions. We will be interested in anyons interacting with a repulsive $\delta$-interaction and subject to an external harmonic potential. A system of $N$ such particles is described, in the first quantized language, by
\bea
H= \sum_{i}^{N} \left( -\frac{\hbar^2}{2m}\frac{\partial^2}{\partial x_{i}^{2}} +\frac{1}{2} m \omega^2 x_i^2 \right) +
2c\sum_{1\leq i<j\leq N}\delta(x_i-x_j).
\label{ham}
\eea
More specifically, we will deal with the Tonks-Girardeau limit $c\to \infty$. 
By taking the distance and energy units as $(\hbar/m\omega)^{1/2}$ and $\hbar\omega/2$ respectively, the single particle eigenstates of the Hamiltonian Eq.~(\ref{ham}) can be written as
\bea
\phi_k(x) = \frac{2^{-k}}{c_k^H} e^{-x^2/2} H_k(x), \qquad k=0,1,2,\ldots,
\eea
where $(c_k^H)^2= 2^{-k} \pi^{1/2} k!$ and $H_k(x)$ is the $k$-th Hermite polynomial.  In case of fermions, it is well-known that the $N$-body ground-state wavefunction is
\bea
\fl
\psi^F_N(x_1,\cdots,x_N)= \frac{1}{\sqrt{N!}} \det [\phi_{j-1}(x_k) ]_{j,k=1,\ldots N} = \non
= \frac{1}{\sqrt{N!} \,C^H_N}  \prod_{l=1}^N e^{-x_l^2/2}  \prod_{1\leq i<j\leq N} (x_i-x_j), \qquad C^H_N=\prod_{k=0}^{N-1}c^H_k ,
\label{wffermion}
\eea
where $\prod_{1\leq i<j\leq N} (x_i-x_j)\equiv \Delta_N(x)$ is a Vandermonde determinant, which appears because  
$2^{-k}H_k(s)$ is a monic polynomial and the determinant is a multilinear alternating form. 
The ground-state wavefunctions of $N$ anyons with statistical parameter $\k$ can be found from here by applying the anyon-fermion mapping \cite{girardeau-06}
\begin{equation}
 \psi^\k_N(x_1,\ldots,x_N)=\left[\prod_{1\leq i < j
      \leq N} A(x_j-x_i) \right]\psi^F_N(x_1,\cdots,x_N),
  \label{AF}
\end{equation}
with
\be
A^\k(x_j-x_i)=\cases{ 
e^{i\pi (1-\k)}& $x_j<x_i$, \cr 
1                  & $x_j>x_i$.
}  
\label{1dbraiding}
\ee
Note that $\psi^F_N(x_1,\ldots,x_N)=\psi^1_N(x_1,\ldots,x_N)$. The relation 
\bea
\fl
 \psi^\k_{N+1}(t,x_1,\ldots, x_N) = \frac{e^{-x^2/2}}{\sqrt{N+1} c^H_N} \prod_{l=1}^N A^\k(x_l-t) (x_l-t)  \psi^\k_N(x_1,\ldots,x_N),
\eea
allows us to express the one-body reduced density matrix of this system as
\bea
\fl \rho^\k_{N+1} (t,t') = (N+1) \int d^Nx \, \bar{\psi}^\k_{N+1} (t,x_1,\ldots,x_N) \psi^\k_{N+1} (t',x_1,\ldots,x_N) \non
= \frac{2^N}{N!} \frac{e^{-t^2/2} e^{-t'^2/2}}{\sqrt{\pi}} \int d^Nx \prod_{l=1}^N \bar{A}^\k(x_l-t) (x_l-t) \non  \times A^\k(x_l-t') (x_l -t')   \left[ \psi^F_{N} (x_1,\ldots  ,x_N )\right]^2
\label{rhoprod}
\eea
where we have used $ \left| \psi^\k_{N} (x_1,\ldots  ,x_N )\right|^2= \left[ \psi^F_{N} (x_1,\ldots  ,x_N )\right]^2$. Plugging   Eq.~(\ref{wffermion}) into Eq.~(\ref{rhoprod}), we see that the latter is suitable for the application  of the identity (see Eq.~(74) in \cite{forrester-02})
\bea
\frac{1}{N!}  \prod_{l=1}^N \int_{-\infty}^\infty dx_l g(x_l) \left(\det [f_{j-1}(x_k) ]_{j,k=1,\ldots N} \right)^2 \non = \det\left[\int_{-\infty}^\infty  ds g(s) f_{j-1}(s) f_{k-1}(s)   \right]_{j,k=1,\ldots N} ,
\eea
from which
\bea
\fl
\rho^\k_{N+1} (t,t') = \frac{2^N}{N!} \frac{e^{-t^2/2} e^{-t'^2/2}}{\sqrt{\pi}}  \det[\int_{-\infty}^\infty  ds \bar{A}^\k(s-t)   (s-t)  \non  \times A^\k(s-t')  (s-t') \phi_j(s) \phi_k(s) ]_{j,k=0,\ldots N-1}  \non
\equiv \frac{2^N}{N!} \frac{e^{-t^2/2} e^{-t'^2/2}}{\sqrt{\pi}}  \det \left[\frac{2^{(i+j)/2}}{2 \sqrt{\pi \Gamma(i) \Gamma(j)} }  a^\k_{jk} (t,t')\right]_{j,k=1,\ldots N} ,
 \label{hankeldet}
\eea
where in the last line we have relabeled  as $i,j \to i+1,j+1$, and employing the multilinearity of the determinant we can write
\bea
\fl
a^\k_{jk}(t,t') =\int_{-\infty}^\infty  ds \bar{A}^\k(s-t) (s-t)  A^\k(s-t')  (s-t') s^{i+j-2} e^{-s^2}, \label{asymbol}
\eea
which is manifestly a Hankel matrix. Let us express the above equation   in terms of (incomplete) gamma functions \cite{forrester-02}. Using the definition Eq.~(\ref{1dbraiding})
\bea
\fl
a^\k_{jk}(t,t') =\int_{-\infty}^\infty  ds (s-t) (s-t') s^{i+j-2} e^{-s^2}  \non
-(1-e^{-i \pi (1-\k)\epsilon(t'-t)}) \epsilon(t'-t) \int_t^{t'}  ds (s-t) (s-t') s^{i+j-2} e^{-s^2} \non
\fl
\equiv f_{j,k}(t,t') 
 -(1-e^{-i \pi (1-\k)\epsilon(t'-t)}) \epsilon(t'-t)  \non \times \left[tt' \mu_{j+k-2}(t,t') -(t+t')\mu_{j+k-1}(t,t') +\mu_{j+k}(t,t')) \right].
 \label{asymbol2}
\eea
In the last step we have introduced the functions
\bea
\fl
f_{j,k}(t,t')= \int_{-\infty}^{\infty} ds\; (s-t) (s-t')  s^{j+k-2} e^{-s^2}
\non
= \cases{
\Gamma\Big({j+k-1\over 2}\Big) tt'
  +\Gamma\Big({j+k+1\over 2}\Big) & \textit{j}+\textit{k} even\cr
-\Gamma\Big({j+k\over 2}\Big)(t+t') &\textit{j}+\textit{k}  odd}
\\
\fl \mu_m(t,t') = \int_t^{t'} ds\; s^m e^{-s^2} 
= {(\epsilon(t'))^{m+1}\over 2}\gamma\Big({m+1\over 2},t'^2\Big)
-{(\epsilon(t))^{m+1}\over 2}\gamma\Big({m+1\over 2},t^2\Big)
\eea
where $\gamma$ is the lower incomplete gamma function, $\gamma(m,x)=\int_0^x ds\, s^{m-1} e^{-s} $.

\subsection{Symmetries} \label{sec:symmdet}

It is important to note the symmetry properties of $\rho^\k_{N+1} (t,t')$. In particular
\begin{itemize}
\item[i)] $\rho^\k_{N+1} (t',t)= \overline{\rho^\k_{N+1}} (t,t')$ (coordinate exchange);
\item[ii)] $\rho^\k_{N+1} (-t,-t')= \overline{\rho^\k_{N+1}} (t,t')$ (center reflection).
\end{itemize}
Looking at Eq.~(\ref{asymbol}) the first property is evident. As for the second one, changing integration variable from  $s$ to $-s$ and recalling that $\bar{A}^\k(-x) A^\k(-x')=A^\k(x) \bar{A}^\k(x')$ we get $a^\k_{jk}(-t,-t')= (-1)^{j+k} \overline{a^\k_{jk}} (t,t')$; the factor $ (-1)^{j+k}$, however, does not change the determinant.
It is also interesting to mention that it holds $\rho^{-\k}_{N+1} (t',t)= \overline{\rho^\k_{N+1}} (t,t')$.

\section{One-body reduced density matrix in the replica approach} \label{sec:replica}

We have  showed that the one-body reduced density matrix can be expressed exactly as a determinant of a Hankel matrix whose coefficients can be written in terms of special functions. This form is particularly suitable for numerical computation and we will provide some examples in Sec.~\ref{sec:numerics}. For increasing $N$, however, the computational cost becomes considerably larger, which is one of the main motivations why it is interesting to seek for an analytic expression that can provide accurate results under well-defined conditions. It has been proved in several cases, including impenetrable bosons \cite{gangardt-04} and anyons \cite{calabrese-08} in a circular geometry and impenetrable bosons in a harmonic potential \cite{gangardt-04}, that by using a replica trick one can find the full asymptotic expansion of $\rho_N(t,t')$ for large $N$ which is valid in the domain $|t-t'|>r_0/N$ and agrees to an excellent degree with the available numerical results; $r_0$ is the typical length scale of the problem, {\it e.g.} the length of the circle in a circular geometry or the Fermi-Thomas radius in a harmonic well. 
Other models have also been studied by means of the same replica approach \cite{other-rep}.
The purpose of this section is to extend the replica method to impenetrable anyons subject to harmonic trapping.

\subsection{Review of the bosonic case} \label{sec:repbos}

In this subsection we briefly review the replica-type calculation of the one-body density matrix in the bosonic case, essentially following the work of Gangardt \cite{gangardt-04} (which was based on a general trick introduced by Kurchan \cite{k-91}). 
Distances are measured in units of half of the Fermi-Thomas radius, $R_{FT}/2=\sqrt{\hbar N/2 m \omega}$. The original units of Eq.~(\ref{ham}) can be recovered by rescaling $x\to (R_{FT}/2) x$; the ones of Eq.~(\ref{hankeldet}) simply by $x\to \sqrt{N/2} \,x$. One-particle eigenstates are now given by (with a little abuse of notation)
\begin{equation}
 \phi_m(x)={1\o b_m} H_m\left( \sqrt{N/ 2} \,x\right) e^{-{N\o 4} x^2} ,\qquad b_m^2=\left(2\pi\o N\right)^{1\o 2} 2^m m!,
\end{equation}
and the $N$-body ground-state wavefunction reads
\begin{equation}
\fl
 \psi^0_N(x_1,\ldots,x_N) = \frac{1}{\sqrt{N!}} \left| \det_{k,l} \phi_{k-1}(x_l) \right| = {1\o \sqrt{S_N(N)} } |\Delta_N(x)| e^{-{N\o 4} \sum_{i=1}^N
x_i^2}
\end{equation}
where $\Delta_N(x)$ is the usual Vandermonde determinant and the normalization constant ${S_N(N)} $ is expressed by the Selberg integral of Hermite
type
\begin{equation}
\fl
 S_N(\lambda)= \int_{-\infty}^{\infty} d^N x \Delta_N^2(x) e^{-{\lambda \o 2} \sum_{i=1}^N x_i^2} = \lambda^{-{N^2\o 2}} (2\pi)^{N\o 2}
\prod_{i=1}^N \Gamma(1+j).
\label{selberg}
\end{equation}
Let us define the replicated average
\begin{equation}
\fl
 Z_m(t_1,\ldots,t_m)= {1\o S_N(N)} \int_{-\infty}^{\infty} d^Ny \Delta_N^2(y) e^{-{N\o 2} \sum_{i=1}^N y_i^2} \prod_{i=1}^N
\prod_{a=1}^m (t_a-y_i).
\end{equation}
Then the one-body reduced density matrix can  be formally obtained by taking the limit $n\to 1/2$:
\begin{eqnarray}
\fl
 \rho_{N+1}^0(t,t') = (N+1) {S_N(N)\o S_{N+1}(N)} e^{-{N\o 4} (t^2 + t'^2)} \lim_{n\to {1\o 2}} Z_{4n}
(\underbrace{t,\ldots,t}_{2n},\underbrace{t',\ldots,t'}_{2n}) \non  \equiv (N+1) {S_N(N)\o S_{N+1}(N)} e^{-{N\o 4} (t^2 + t'^2)} \lim_{n\to {1\o
2}}
Z_{4n}(t,t').
\label{rhorep}
\end{eqnarray}
$Z_{4n}(t,t')$ is more conveniently expressed after a duality transformation (see \cite{gangardt-04} and references therein)
\begin{eqnarray}
\fl
 Z_{4n}(t,t') = {1\o S^2_{2n}(N)} \int_{-\infty}^{\infty} d^{2n}x d^{2n}x' \Delta_{2n}^2(x) \Delta_{2n}^2(x')  \non \times {\prod_{a,a'=1}^{2n}
(x_a-x'_{a'}) \o [i(t-t')]^{4n^2}} e^{-N\sum_{a=1}^{2n} S(x_a,t)} e^{-N\sum_{a'=1}^{2n} S(x'_{a'},t')},
\label{dualz}
\end{eqnarray}
where we introduced the  ``action''
\begin{equation}
 S(x,t)={(x-it)^2 \o 2} -\log x +{\pi i\o 2}.
 \label{action}
\end{equation}
The saddle points of (\ref{action}) are given by
\begin{equation}
 x_{\pm}= {it\o 2} \pm \sqrt{1-{t^2 \o 4}} = \pm e^{\pm i \phi}, \qquad \sin \phi={t\o 2}, \label{phidef}
\end{equation}
for $t \in ]-2,2[$. At these points
\begin{eqnarray}
 S(x_{\pm},t) &\equiv & S_\pm = e^{\mp 2i\phi} \mp i\phi \pm {\pi i\o2}, \\
S''(x_{\pm},t) &\equiv & \sigma_\pm =2 e^{\mp 2i\phi} \cos \phi.
\end{eqnarray}
It is convenient to define the two functions 
\begin{eqnarray}
 \Theta(t) &=& 2\phi + \sin 2\phi + \pi = 2 \pi \int_{-2}^t \rho(s)\, ds \label{thetadef},\\
 \rho(t) &=& {1\o \pi} \cos \phi = {1\o \pi} \sqrt{1- {t^2 \o 4}} \label{rhodef}.
\end{eqnarray}
The latter  is the well-known Wigner semi-circle law for the mean density of particles in the large $N$ limit. We will use primed symbols, namely $x_\pm',\phi',\Theta',S_\pm',\sigma_\pm'$, for functions of $t'$ (as opposed to $t$).
We will be interested in the regime in which the saddle points are well separated: on the one hand $|t-t'|$ must be of order of the cloud size, namely $|t-t'|= O(1)$ (or $|t-t'|\sim  R_{FT}\sim \sqrt{N}$ in the old units); on the other hand $t,t'$ must be sufficiently far from the edges of the cloud, that is $||t|-R_{FT} |=O(1)$ and similarly for $t'$.
  In order to take into account all the saddle points of Eq.~(\ref{dualz}) we must consider all the possible ways to distribute the variables between the neighborhoods of $x_-$ and $x_+$, namely
\begin{eqnarray}
x_a= x_- + \xi_a/\sqrt{N} \qquad a=1,\ldots, l \non
x_b= x_+ + \xi_b/\sqrt{N} \qquad b=l+1,\ldots, 2n \non
x_{a'}= x_-' + \xi_{a'}/\sqrt{N} \qquad a'=1,\ldots, l' \non
x_{b'}= x_+' + \xi_{b'}/\sqrt{N} \qquad b'=l' +1,\ldots, 2n.
\end{eqnarray} 
When  $l,l'\in \{0,2n\}$  the replica symmetry is preserved, whereas  all the other cases are symmetry-breaking.   The Vandermonde determinants in Eq.~(\ref{dualz})  vanish at the saddle points and are expanded in their vicinity as
\begin{eqnarray}
\fl
\Delta_{2n}^2(x)  = \left(\frac{1}{\sqrt{N}}\right)^{l(l-1)+(2n-l)(2n-l-1)} (x_- - x_+)^{2l(2n-l)} \Delta_{l}^2(\xi_a) \Delta_{2n-l}^2(\xi_b) ,
\end{eqnarray}
and similarly for $\Delta_{2n}^2(x') $. The double product in Eq.~(\ref{dualz}) evaluated at a saddle point is
\begin{eqnarray}
\fl
  \prod_{a=1}^{2n}\prod_{a'=1}^{2n} (x_a-x'_{a'}) =&  i^{4n^2}\, i^{2n(l+l'-2n)}
  \left|t-t'\right|^{2n^2} \non 
 & \times  \left[\frac{\cos^2\frac{\phi+\phi'}{2}}
  {\sin^2\frac{\phi-\phi'}{2}}\right]^{(l-n)(l'-n)}
  e^{-2in (l-n)\phi + 2in(l'-n)\phi'}.
\end{eqnarray}
We can now apply the saddle point method to Eq.~(\ref{dualz}) and calculate both the contribution from the stationary value of the action and the one from the fluctuations; the latter can be found by using the Selberg integral of Eq.~(\ref{selberg}). Combining all together we obtain
\begin{eqnarray}
\fl
  Z_{4n}(t,t') = 
  \left|t-t'\right|^{-2n^2}  
  \sum_{l=0}^{2n} \sum_{l'=0}^{2n}  (-1)^{n(l+l'-2n)} N^{l(2n-l)+l'(2n-l')}  \left[\frac{\cos\frac{\phi+\phi'}{2}}
  {\sin\frac{\phi-\phi'}{2}}\right]^{2(l-n)(l'-n)} \non
   \times F^l_{2n}  \frac{ (x_+-x_-)^{2l(2n-l)}}
  {\left(\sqrt{\sigma_-}\right)^{l^2}\left(\sqrt{\sigma_+}\right)^{(2n-l)^2}}
  e^{-NlS_--N(2n-l)S_+ -2in (l-n)\phi} \non
  \times F^{l'}_{2n}  \frac{ (x_+'-x_-')^{2l' (2n-l')}}
  {\left(\sqrt{\sigma_-'}\right)^{l'^2}\left(\sqrt{\sigma_+'}\right)^{(2n-l')^2}} 
  e^{-Nl'S_-' -N(2n-l')S_+' +2in(l'-n)\phi'}.
  \label{zsaddle}
\end{eqnarray}
The $F$-symbols $F_{2n}^l $ are defined as
\begin{eqnarray}
  F_{2n}^l &=& {2n \choose l} \frac{
    \prod_{a=1}^l\Gamma(a+1)\;\prod_{b=1}^{2n-l}\Gamma(b+1)} {
    \prod_{c=1}^{2n} \Gamma(c+1)} \non
    &=& \prod_{a=1}^l\frac{\Gamma(a)}
    {\Gamma(2n+1-a)} 
    = \frac{G(l+1)G(2n-l+1)}{G(2n+1)},
  \label{eq:fsymbol}
\end{eqnarray}
where $G(x)$ is the Barnes $G$-function. For integer $n$ we can extend the double sum in Eq.~(\ref{zsaddle}) to all integers because in this case $ F_{2n}^l$ vanishes for $l<0$ and $l>2n$. This step might appear insignificant at this stage, but is actually important when we take the limit $n\to 1/2$ and break the replica symmetry, because in that case $F_{1}^l$ is non-zero for any $l$ and the two sums in the replicated average are genuinely infinite. 
At this point we change the summation variables in Eq.~(\ref{zsaddle}) to 
\be
m=l-n, \qquad  m'=l-n. \label{repsbbos} 
\ee
This is actually a crucial step, that will allow to obtain the correct analytic continuation, or the correct replica symmetry breaking,  for the bosonic statistics. Using Eqs. (\ref{phidef})-(\ref{rhodef}), after quite a lot of algebra one arrives to 
\bea
\fl
Z_{4n}(t,t') = (2\pi N)^{2n^2} \, e^{-Nn(2-(t^2+t'^2)/2)} \,
\frac{[\rho(t)\rho(t')]^{n^2}}{|t-t'|^{2n^2}} \non
\times \sum_{m=-\infty}^{\infty} \sum_{m'=-\infty}^{\infty}  \left\vert
{\cos{\phi+\phi' \over 2} \over \sin{\phi-\phi'\over 2}} \right\vert^{2mm'}
{(-1)^{nm} F_{2n}^{n+m} \over [8N \pi^3 \rho^3(t)]^{m^2}} e^{-i Nm \Theta
-4inm\phi} \, \non \times  {(-1)^{nm'} F_{2n}^{n+m'} \over [8N \pi^3 \rho^3(t')]^{m'^2}}
e^{i Nm' \Theta' +4inm'\phi'}. \label{zbosonic}
\eea
We are now in the position to take the limit as indicated in Eq.~(\ref{rhorep}). Also, noting that
\be
\fl
\frac{S_N(N)}{S_{N+1}(N)} = { N^{N+{1\over 2}} \over (2\pi)^{1\over 2}
\Gamma(N+2)}= { N^{N+{1\over 2}} \over (2\pi)^{1\over 2} (N+1)!} = {e^N
\over 2\pi N} \left[ 1+O\left(\frac{1}{N} \right) \right],
\label{largen}
\ee
we finally arrive to
\begin{eqnarray}
\fl
\rho_{N+1}^0(t,t') 
 = \left(\frac{2\pi}{N} \right)^{{1\over 2}} \, \frac{[\rho(t)\rho(t')]^{1\over
4}}{|t-t'|^{1\over 2}} \sum_{m=-\infty}^{\infty} \sum_{m'=-\infty}^{\infty}
(-1)^{(m+m')/2} \,\non
\times   F_{1}^{{1\over 2}+m} \, F_{1}^{{1\over
2}+m'}  \left\vert {\cos{\phi+\phi' \over 2} \over \sin{\phi-\phi'\over 2}}
\right\vert^{2 m m'}  
  {e^{-i m (N \Theta
+2\phi) +i m' (N \Theta' +2\phi')}
\over [8N \pi^3 \rho^3(t)]^{m^2} [8N \pi^3 \rho^3(t')]^{m'^2}}.
\label{rhomacrbos}
\end{eqnarray}
It is important to note that the zero mode ($m,m'=0$) selected by the choice in Eq.~(\ref{repsbbos}), which gives the leading order in $1/N$, coincides exactly with the asymptotic formula of \cite{forrester-02}. This confirms that correctness of the analytic continuation performed above.
The zero mode also coincides with the asymptotic prediction from the so called trap-size scaling \cite{tss}.

\subsection{Anyons: macroscopic limit}

For a system of particles with generalized anyonic statistics in a harmonic potential we will employ the replica method of the previous subsection with the modified prescriptions found and applied to the case of circular geometry in Ref. \cite{calabrese-08}. 

First of all let us assume that the statistical parameter is a rational number, $\k=q/p$. This is only part of the replica construction and does not pose any restriction to the final results, which will be valid for any real $\kappa$ between 0 and 1.
The one-particle density matrix for anyons is defined through the replicated average as 
\begin{equation}
 \rho_{N+1}^\k(t,t')= (N+1) \frac{S_N(N)}{S_{N+1}(N)} e^{-\frac{N}{4} (t^2+t'^2)}
\lim_{2np\to 1} Z^q_{4np} (t,t'),
\end{equation}
where $q$ indicates which branch should be taken in order to recover the appropriate analytic continuation as it will be clear below.
In order to find the correct replicated average we do not need to repeat all the steps of Sec.~\ref{sec:repbos}, but we can just start from Eq.~(\ref{zsaddle}) with the replacement $n\to np$. Now the correct change of summation variables is given by
\be
m+nq=l-np, \qquad  m'+nq=l-np. \label{repsbany} 
\ee
%
which leads to
\begin{eqnarray}
\fl
Z^q_{4np}(t,t') = (2\pi N)^{2(np)^2} \, e^{-Nnp(2-(t^2+t'^2)/2)} \,
\frac{[\rho(t)\rho(t')]^{(np)^2}}{|t-t'|^{2(np)^2}} 
\non \times 
\sum_{m=-\infty}^{\infty} \sum_{m'=-\infty}^{\infty}  \left\vert
{\cos{\phi+\phi' \over 2} \over \sin{\phi-\phi'\over 2}}
\right\vert^{2(nq+m)(nq+m')} \non 
\times {(-1)^{np(nq+m)} F_{2np}^{np+nq+m} \over [8N \pi^3
\rho^3(t)]^{(nq+m)^2}} e^{-i N(nq+m) \Theta -4inp(nq+m)\phi} \non
\times {(-1)^{np(nq+m')} F_{2np}^{np+nq+m'} \over [8N \pi^3
\rho^3(t')]^{(nq+m')^2}} e^{i N(nq+m') \Theta' +4inp(nq+m')\phi'}. \label{zq}
\end{eqnarray}
Now we can analytically continue by taking the limit $2np\to 1$
\begin{eqnarray}
\fl
\lim_{2np\to 1} Z^q_{4np} (t,t') &=& (2\pi N)^{1\over 2} \, e^{-N+{N\over
4}(t^2+t'^2)} \, \frac{[\rho(t)\rho(t')]^{1\over 4}}{|t-t'|^{1\over 2}} 
\non & & \times  \sum_{m=-\infty}^{\infty} \sum_{m'=-\infty}^{\infty}  \left\vert
{\cos{\phi+\phi' \over 2} \over \sin{\phi-\phi'\over 2}}
\right\vert^{2({\k\over2}+m)({\k \over 2}+m')} \non 
&&\times  {(-1)^{{1\over 2}({\k\over 2}+m)}
F_{1}^{{1\over 2}+{\k \over 2}+m} \over [8N \pi^3 \rho^3(t)]^{({\k \over
2}+m)^2}} e^{-i N({\k\over 2}+m) \Theta -2i({\k\over 2}+m)\phi} \non
&&\times {(-1)^{{1\over 2}({\k\over 2}+m')} F_{1}^{{1\over 2}+{\k\over 2}+m'}
\over [8N \pi^3 \rho^3(t')]^{({\k\over 2}+m')^2}} e^{i N({\k\over 2}+m')
\Theta' +2i({\k\over 2}+m')\phi'}.
\end{eqnarray}
Recalling Eq.~(\ref{largen}) we end up with
\begin{eqnarray}
\fl
\rho_{N+1}^\k(t,t') 
 = & \left(\frac{N}{2\pi}\right)^{\frac{1}{2}} \, \frac{[\rho(t)\rho(t')]^{1\over
4}}{|t-t'|^{1\over 2}} \sum_{m=-\infty}^{\infty} \sum_{m'=-\infty}^{\infty}
(-1)^{(\k+m+m')/2} \, F_{1}^{{1\over 2}+{\k\over 2}+m} \, F_{1}^{{1\over
2}+{\k\over 2}+m'} \non\fl
&\times  \left\vert {\cos{\phi+\phi' \over 2} \over \sin{\phi-\phi'\over 2}}
\right\vert^{2({\k\over2}+m)({\k \over 2}+m')}  {e^{-i ({\k\over 2}+m)(N \Theta
+2\phi) +i ({\k\over 2}+m') (N \Theta' +2\phi')}
\over [8N \pi^3 \rho^3(t)]^{({\k \over 2}+m)^2} [8N \pi^3 \rho^3(t')]^{({\k\over
2}+m')^2}}.
\label{rhomacr}
\end{eqnarray}

\subsection{Anyons: mesoscopic limit} \label{sec:anyonmes}

In the so-called mesoscopic regime $t,t'$ are comparable with the mean inter-particle distance $1/N\rho(t)$ but  such that $|t-t'|\gg 1/N$, because this implies that there is still a large number of particle between them and we can safely focus on the asymptotic behavior of correlations. We concentrate on the region around the center
of the potential, setting $t+t'=0$, and define the scaling variable $t-t'=x/N$.
For small $t,t'$
$$
\left\vert {\cos{\phi+\phi' \over 2} \over \sin{\phi-\phi'\over 2}}
\right\vert^{2({nq}+m)({nq}+m')} \simeq \left(4N\over
x\right)^{2({nq}+m)({nq}+m')}.
$$
Combining with the powers of $N$ in (\ref{zq}) we get a factor of
$N^{-(m-m')^2}$, meaning that only diagonal terms $m=m'$ give the leading
contribution (other terms are at least $1/N$ smaller). Also, as discussed in
\cite{gangardt-04}, the two limits $N\to \infty$ and $t-t'\to 0$ commute, which comes from
the details of the saddle point integration; therefore, this result is
insensitive to the introduction of a generalized statistical parameter
(technically to the choice of branch of the replicated average) and we can
safely use (\ref{zq}) with $\rho(t),\rho(t') \simeq 1/\pi$,
$\Theta-\Theta',4\phi-4\phi' \simeq 2x/N$:
\begin{eqnarray}
\fl
Z^q_{4np}(x) = \left(2N^2\over x\right)^{2(np)^2} e^{-Nnp(2-(t^2+t'^2/2))} \non
 \times \sum_{m=-\infty}^{\infty} (-1)^{2np(nq+m)} \left[F_{2np}^{np+nq+m}\right]^2 {
e^{-2i(nq+m)(1+{np/N})x}\over (2x)^{2(nq+m)^2}},
\end{eqnarray}
\bea
\fl
\lim_{2np\to 1} Z^q_{4np}(x) = \left(2N^2\over x\right)^{1\over 2}
e^{-N-{N\over 4}(t^2+t'^2)} \non 
\times \sum_{m=-\infty}^{\infty} (-1)^{{\k\over 2}+m}
\left[F_{1}^{{1\over 2}+{\k\over 2}+m}\right]^2 { e^{-2i({\k\over
2}+m)(1+{1/2N})x}\over (2x)^{2({\k\over 2}+m)^2}},
\eea
\bea
\fl
\rho_{N+1}^\k(x)= {N\over \sqrt{2} \pi \, x^{1/2}} \sum_{m=-\infty}^{\infty}
(-1)^{{\k\over 2}+m} \left[F_{1}^{{1\over 2}+{\k\over 2}+m}\right]^2 {
e^{-2i({\k\over 2}+m)(1+{1/2N})x}\over (2x)^{2({\k\over 2}+m)^2}}.
\label{rhomes}
\eea
This expansion has exactly the same form as the one obtained in \cite{calabrese-07} from the bosonization approach to a system of anyons on a circle when the Luttinger parameter equals 1 (Tonks-Girardeau limit). The different geometry is not an issue here, because in the mesoscopic regime near the center of the cloud the curvature of the trap plays a minor role. The theoretically interesting aspect concerns the exponents of the $1/N$ series, which the replica method generates correctly. Even more striking is the fact that the same method produces also all the correct coefficients in Eq.~(\ref{rhomacr}) and Eq.~(\ref{rhomes}), as confirmed below in Sec.~\ref{sec:numerics}.

\subsection{Symmetries}

Let us check that $\rho^\k_{N+1}(t,t')$ in Eq.~(\ref{rhomacr}) correctly possesses the symmetry properties identified in Sec.~\ref{sec:symmdet}. The behavior under coordinate exchange, that is $\rho^\k_{N+1} (t',t)= \overline{\rho^\k_{N+1}} (t,t')$, is evident. As for the center inversion, we need to consider that from the definitions Eqs.~(\ref{phidef}), (\ref{thetadef}) and (\ref{rhodef}) we have $\phi(-t)=-\phi(t)$, $\Theta(-t)=-\Theta(t)+2\pi$ and $\rho(-t)=\rho(t)$; the last numerator in Eq.~(\ref{rhomacr}) is then transformed to its complex conjugate up to a factor $e^{2N\pi i(m'-m)}=1$, which is enough to prove $\rho^\k_{N+1} (-t,-t')= \overline{\rho^\k_{N+1}} (t,t')$.


\section{Numerics and comparison with the replica method} \label{sec:numerics}

In this section we  present a sample calculation of $\rho^\k_{N+1}(t,t')$ using the exact representation in Eq.~(\ref{hankeldet}) and we carefully compare the numerical results with the asymptotic expansion Eq.~(\ref{rhomacr}). The unit of length is set to  $R_{FT}/2$ as in Sec.~\ref{sec:replica}. We will see that the first few terms in the expansion are sufficient to realize an excellent approximation in the regions where the saddle point treatment is justified (see discussion in Sec.~\ref{sec:repbos}).

\begin{figure}[tb]
\begin{center}
\includegraphics[width=\textwidth]{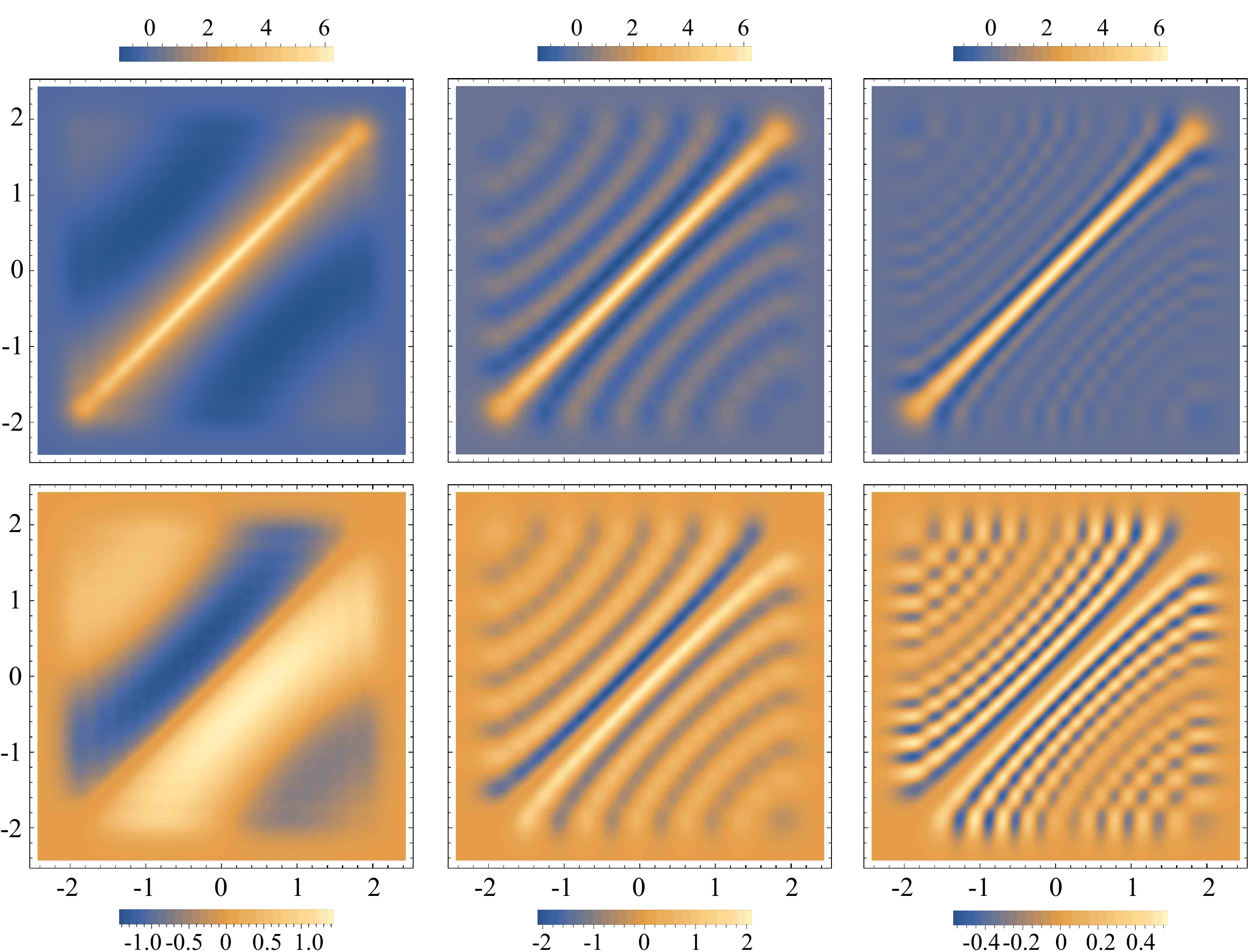}
\end{center}
\caption{Real (upper row) and imaginary (lower row) part of $\rho_{20}^{\k}(t,t')$, $\k=0.1,0.5,0.9$ (left to right), calculated from the exact representation in  Eq.~(\ref{hankeldet}). Lengths are in units of $R_{FT}/2$.}
\label{fig:xyall}
\end{figure}

In Fig.~\ref{fig:xyall} we display the full one-body reduced density matrix for $N=20$ and $\k=0.1,0.5,0.9$ as a density plot over the $[t,t']$-plane. For $\kappa\neq 0,1$ a non-zero imaginary part develops, which is a general feature of generalized statistics independently of the geometry; a striking consequence of this is that  the momentum distribution function,  defined by $n^\k_N(k)= (1/2\pi) \int dt\int dt' e^{i k (t'-t)} \rho^\k_N (t,t')$, is asymmetric for reflections about $k=0$, as already known for circular geometry \cite{santachiara-07,santachiara-08}. 
Because of the harmonic trapping,  $\rho^\k_{N+1}(t,t')$ vanishes very quickly for $|t|,|t'| >2$, namely outside of the Fermi-Thomas radius, even for relatively small $N$. Also, the symmetry properties identified in Sec.~\ref{sec:symmdet} are manifest in the actual calculation.

\begin{figure}[tb]
\begin{center}
\includegraphics[width=0.49\textwidth]{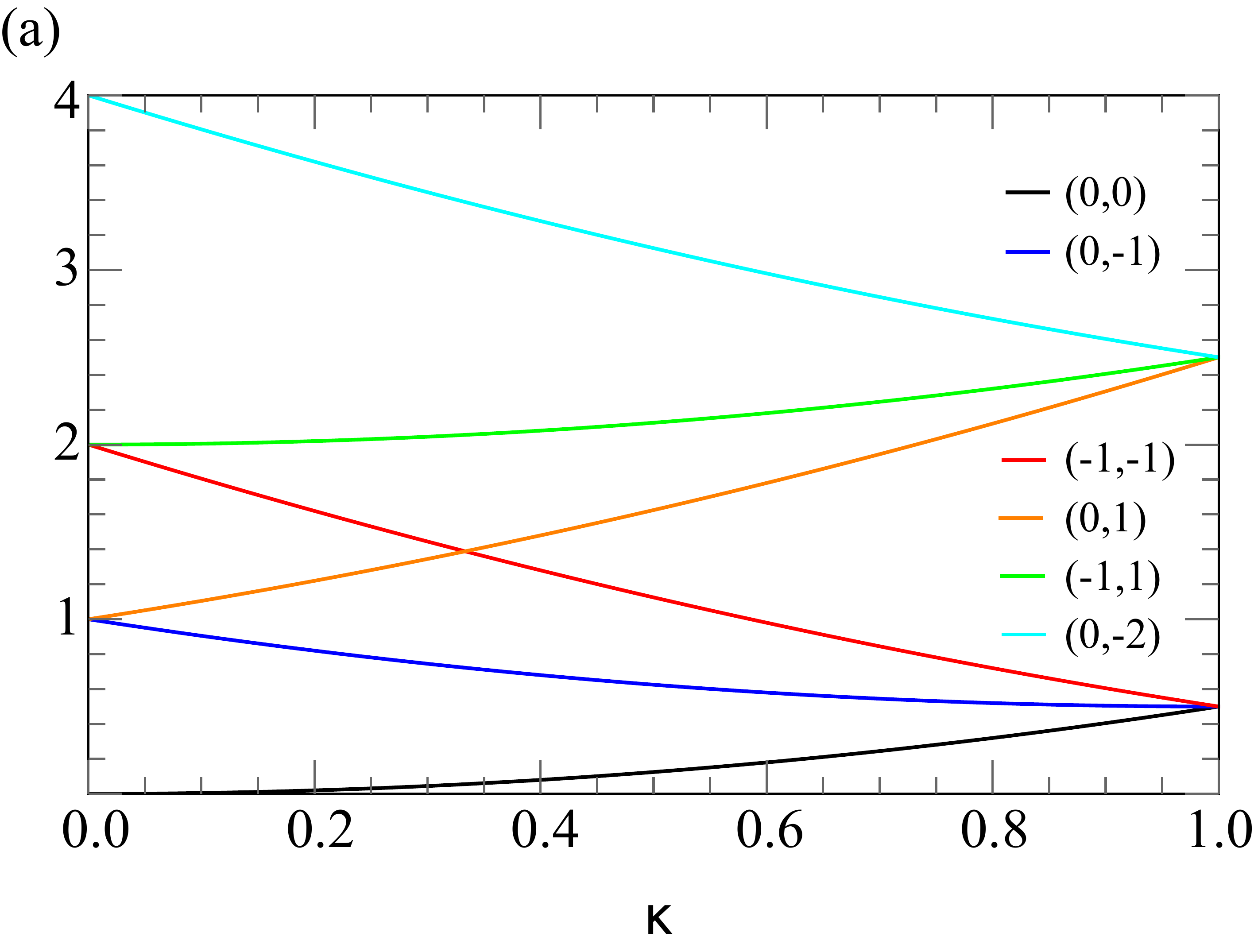}
\includegraphics[width=0.49\textwidth]{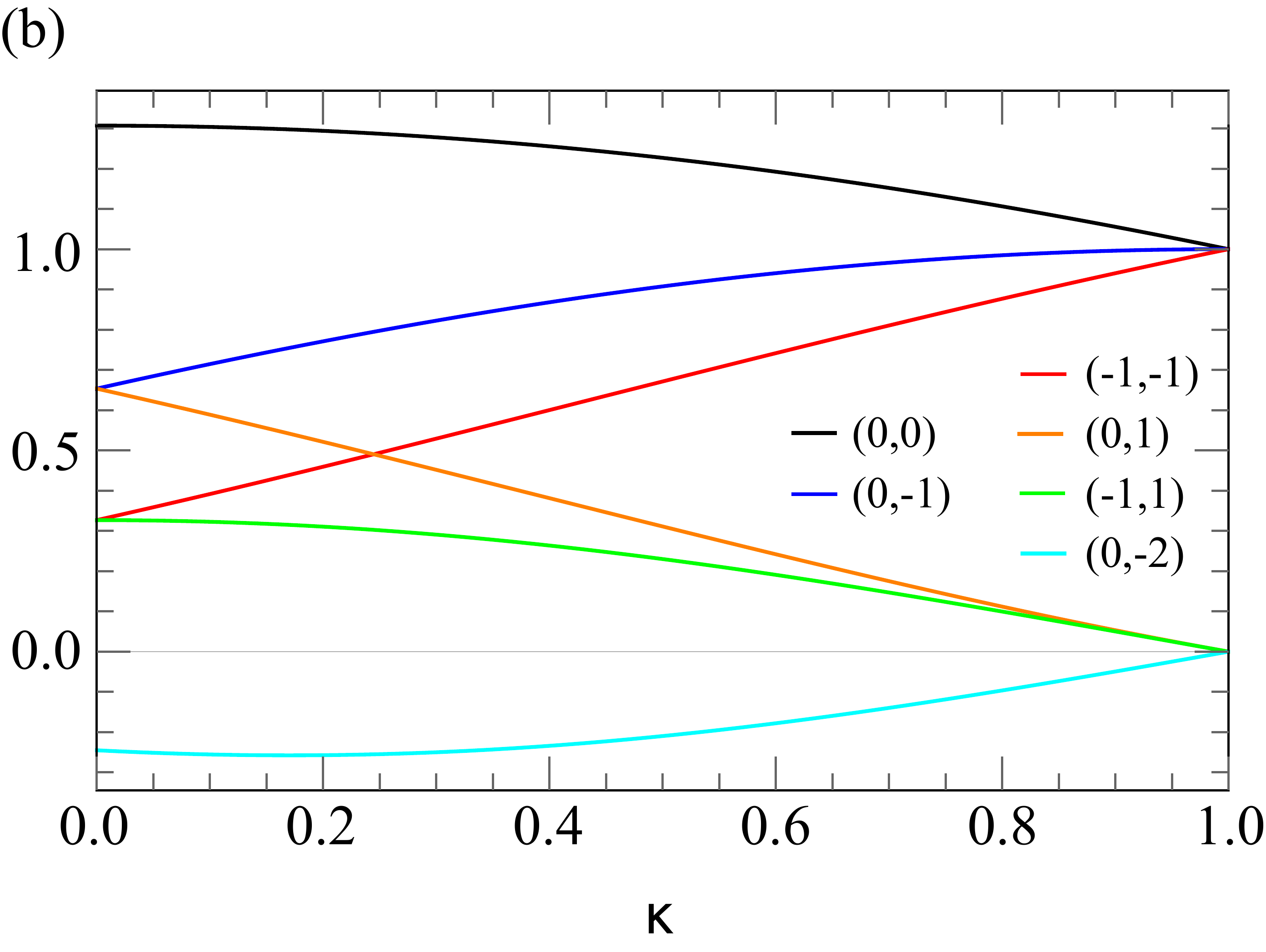}
\end{center}
\caption{(a)  Exponent of $1/N$ of the most relevant terms in the double sum in  Eq.~(\ref{rhomacr}), that is $(\k/2+m)^2 +(\k/2+m')^2$, as a function of the statistical parameter $\k$. (b) The coefficient $F_{1}^{{1\over 2}+{\k\over 2}+m}  F_{1}^{{1\over 2}+{\k\over 2}+m'}$ of the same terms.}
\label{fig:exponents}
\end{figure}

In order to make a comparison with the replica method, we first need to observe the structure of Eq.~(\ref{rhomacr}). Each term in the double sum has an amplitude that scales with a certain  power of $1/N$, namely $(\k/2+m)^2 +(\k/2+m')^2$, and an oscillatory factor (complex exponential) with a characteristic frequency. In particular, the $(m,m')=(0,0)$ term, or zero mode, is always the one with the largest amplitudes and the slowest oscillations, while higher terms are suppressed by a non-integer power of $1/N$ and oscillate faster (except in the limiting case $\k=1$). Since the power series depends on $\k$, one must be careful in choosing  an appropriate truncation of the double sum. In Fig.~\ref{fig:exponents} we plot the  powers of $1/N$ corresponding to the first few terms as a function of $\k$; clearly the term $(m',m)$ shares the same power with the term $(m,m')$ and the two must be always considered together. It is easy to see that for any $0<\k<1$ the next-to-leading term is given by $(0,-1)$ and $(-1,0)$, which have to be taken into account when we want to refine the first approximation given by the zero mode. On the contrary, the terms to be chosen at the next level depend on $\k$: for low $\k$, that is $\k\lesssim 0.2$, the $(0,1)$ and $(1,0)$ terms must be added; for high $\k$, that is $\k\gtrsim 0.4$, the $(-1,-1)$  is more relevant instead; finally, for intermediate $\k$ both must be included to have a consistent truncation because they are essentially of the same order  (note that the corresponding powers of $1/N$ cross at $\k=1/3$). In Fig.~\ref{fig:exponents}(b) the behavior of the first few $F$-symbols is reported to make sure that the previous considerations are not affected by any singular behavior of the numerical coefficients. 

We define a truncation of Eq.~(\ref{rhomacr}) by $\widetilde{\rho}^\k_{N+1}(t,t';D) $, where $D$ is a certain subset of indexes $(m,m')$. In particular, given the above observations, the relevant subsets will be
\begin{eqnarray}
D_0= \{(0,0) \} \non
D_1= \{ (0,0), (0,-1),(-1,0) \} \non
D_l=  \{ (0,0), (0,-1),(-1,0) , (0,1),(1,0) \} \non
D_h=  \{ (0,0), (0,-1),(-1,0) , (-1,-1)\} \non
D_m=  \{ (0,0), (0,-1),(-1,0) , (0,1),(1,0), (-1,-1)\} 
\label{truncation}
\end{eqnarray}
While it is possible to include further terms in the truncation without much effort, it is only of relative interest to do so both for theoretical and practical reasons. On the one hand, one must keep in mind, as shown in \cite{gangardt-04} for the bosonic case,  that each term in Eq.~(\ref{rhomacr}) would acquire a whole series of corrections from a standard perturbation theory around the saddle points Eq.~(\ref{phidef}); the first perturbative  corrections of the first terms (specifically $(0,0), (0,-1),(-1,0)$), although suppressed by a factor $1/N$, will set the actual limitation of the approximation, rather than the higher terms in the double sum. On the other hand, we will see that the truncations introduced in Eq.~(\ref{truncation}) are sufficient to achieve a precision of order $10^{-2}$ (except near the edges or the $t=t'$ line). 

Let us analyze the relative difference 
\be
\frac{\widetilde{\rho}_{N}^{\k}(t,t';D)}{\rho_{N}^{\k}(t,t')}-1
\ee
between the density matrix calculated numerically from Eq.~(\ref{hankeldet}) and  the various truncations of the asymptotic expansion obtained with the replica method. For sake of clearness we focus only on one direction in the two dimensional plane, namely $(0,t)$; furthermore, by exploiting the symmetries we can restrict ourselves to $t>0$. Other choices do not show any qualitative difference in the analysis.

\begin{figure}[tb]
\begin{center}
\includegraphics[width=0.49\textwidth]{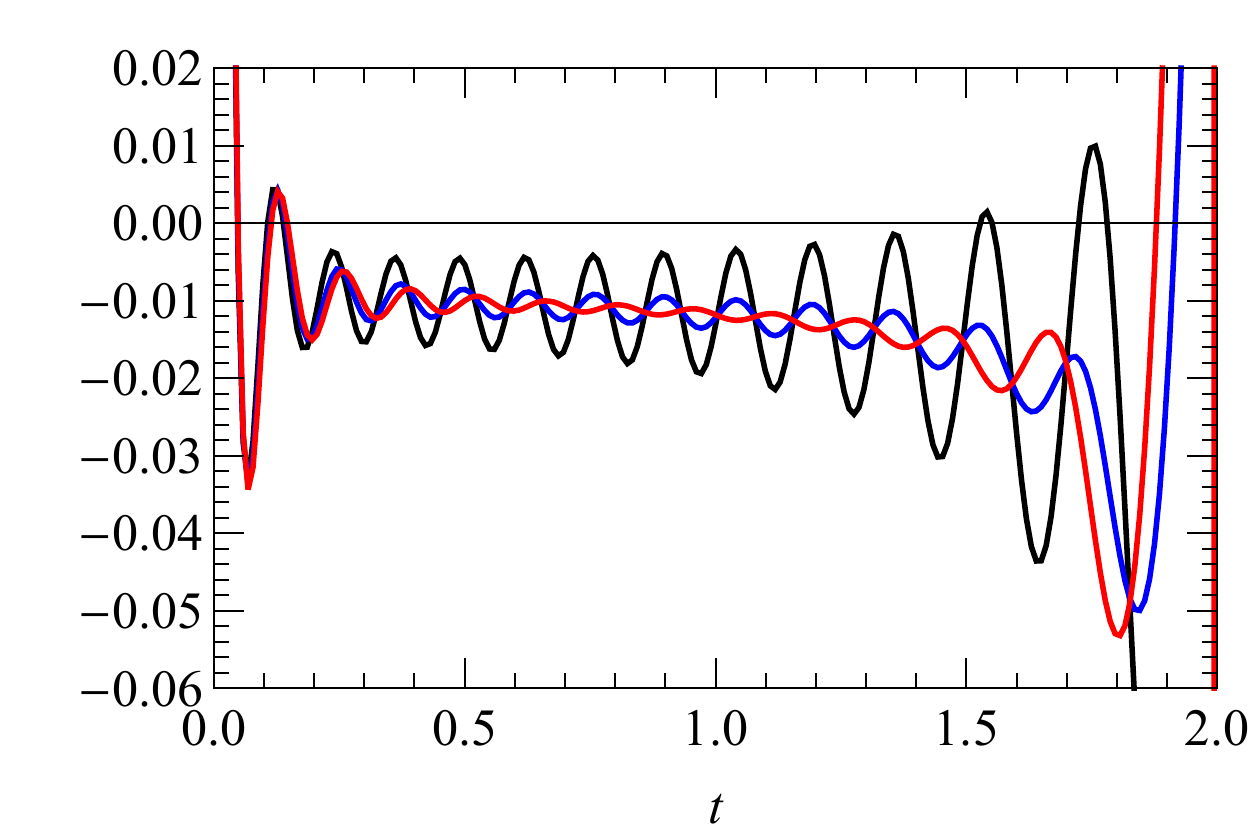}
\includegraphics[width=0.49\textwidth]{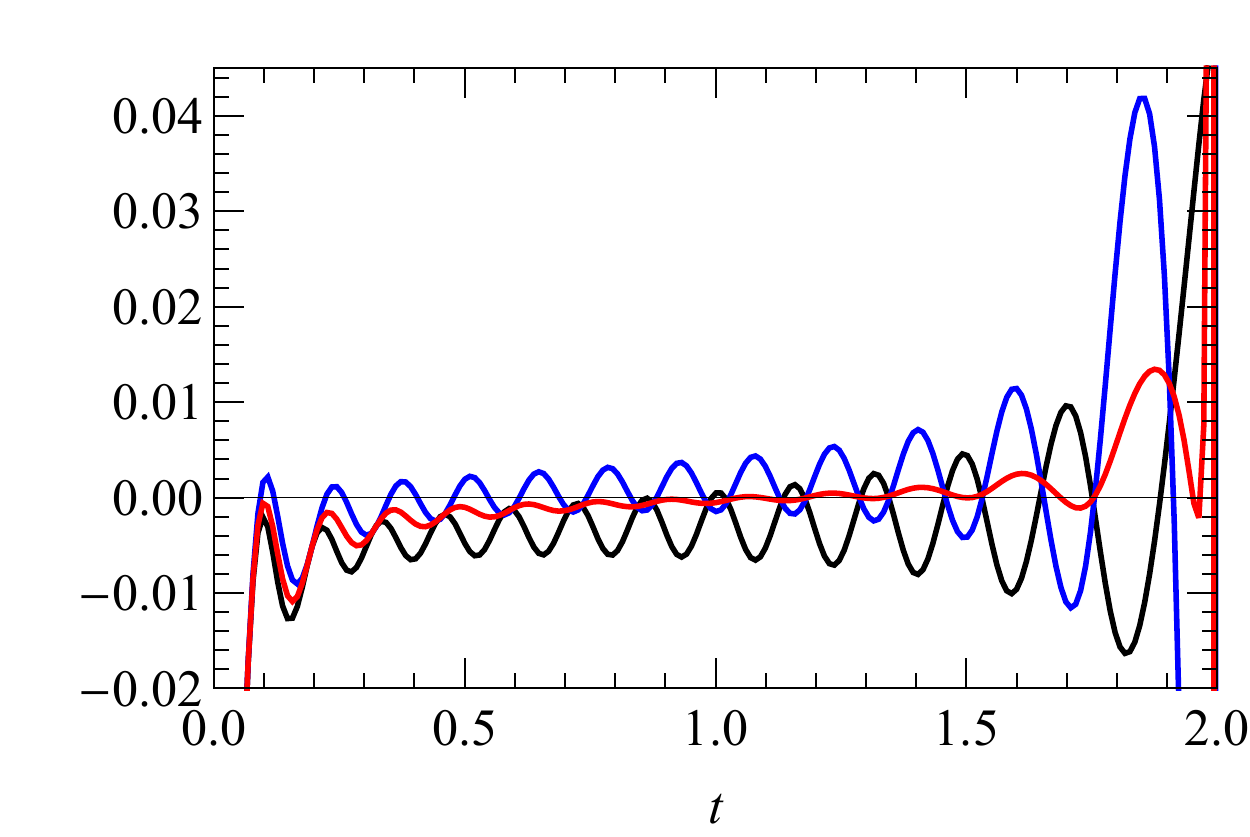}
\end{center}
\caption{Real and imaginary part of $\widetilde{\rho}_{25}^{0.1}(0,t;D) / \rho_{25}^{0.1}(0,t) - 1$ for $D=D_0$ (black), $D=D_1$ (blue) and $D=D_l$ (red).}
\label{fig:k01diff}
\end{figure}

In Fig.~\ref{fig:k01diff} we consider the case of $\k=0.1$ for $N=25$. The zero mode itself already provides a very good approximation; this fact is expected for low $\k$ and visible also in circular geometry \cite{calabrese-08}. Adding the next-to-leading term, namely considering $D_1$, makes evident the preliminary discussion made above. In fact, oscillation are sensibly suppressed, as it appears especially in the real part, but the average does not benefit very much; this is because an improvement in the average actually requires perturbative corrections to the zero mode. By proceeding to the more refined truncation $D_l$ (appropriate for this value of $\k$) we note a further improvement in the oscillations, particularly in the imaginary part, and overall a very good agreement which stays within $0.02$ unless $t$ is very close to the edges.

\begin{figure}[tb]
\begin{center}
\includegraphics[width=0.49\textwidth]{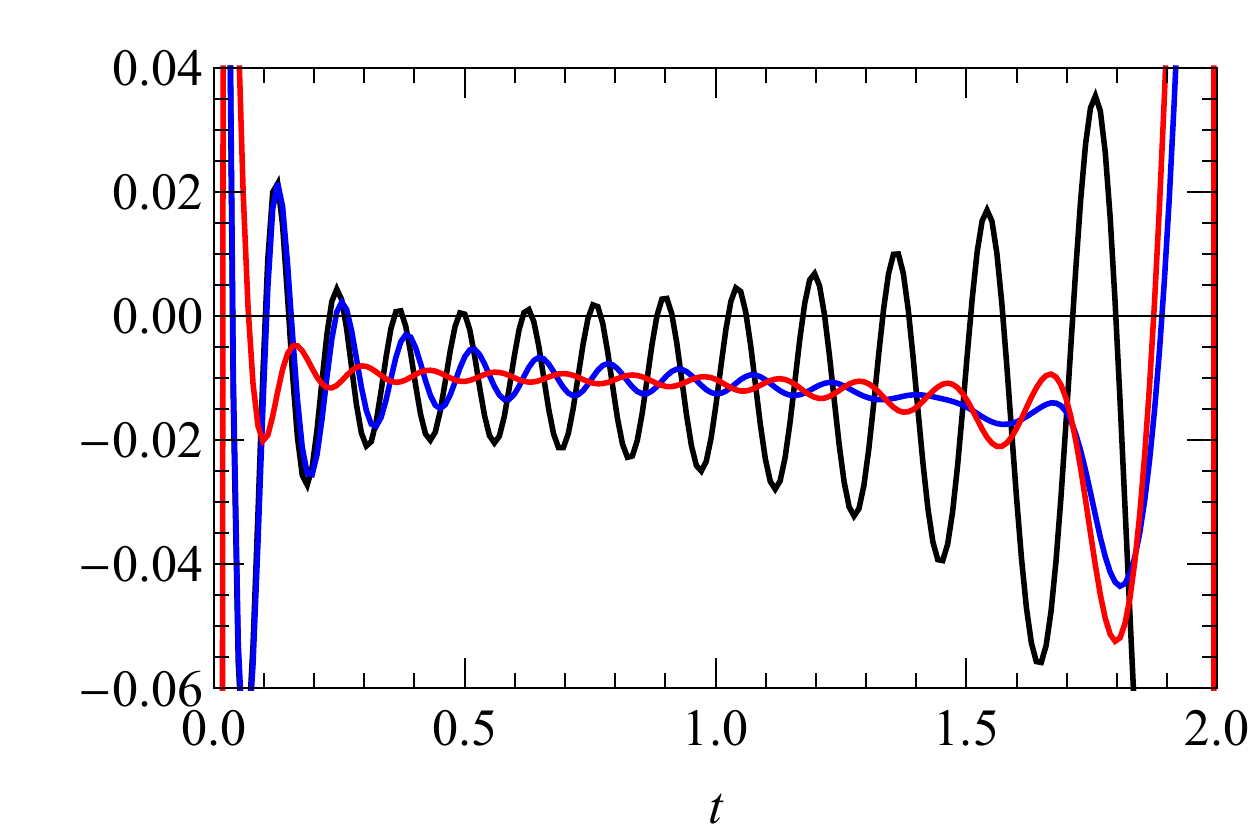}
\includegraphics[width=0.49\textwidth]{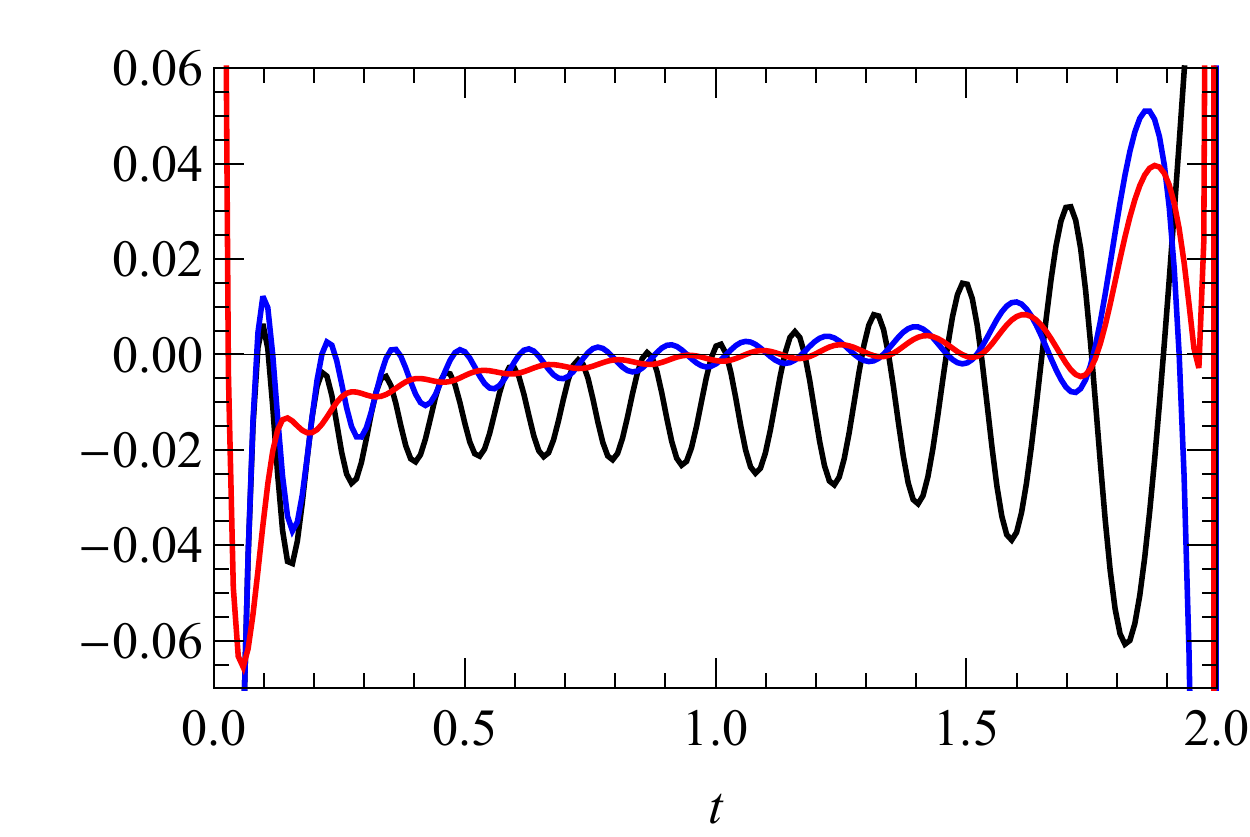}
\end{center}
\caption{Real and imaginary part of $\widetilde{\rho}_{25}^{0.3}(0,t;D) / \rho_{25}^{0.3}(0,t) - 1$ for $D=D_0$ (black), $D=D_1$ (blue) and $D=D_m$ (red).}
\label{fig:k03diff}
\end{figure}

In Fig.~\ref{fig:k03diff} we present the analysis for $\k=0.3$. In this case the zero mode gives a slightly worse, but still remarkable, approximation, which is explained by the fact that its importance and that of the next-to-leading term are getting closer as $\k$ increases (see Fig.~\ref{fig:exponents}). The $D_1$ truncation, instead, is to some extent  more efficient than in the $\k=0.1$ case, which is due to the higher terms being comparatively more suppressed. It is however important to  take the truncation $D_m$, consistent with this intermediate value of $\k$, to further get rid of oscillations, especially in the region of small $t$ (close to $t\sim 1/\sqrt{N}$) where the saddle point treatment is only marginally valid.

\begin{figure}[tb]
\begin{center}
\includegraphics[width=0.49\textwidth]{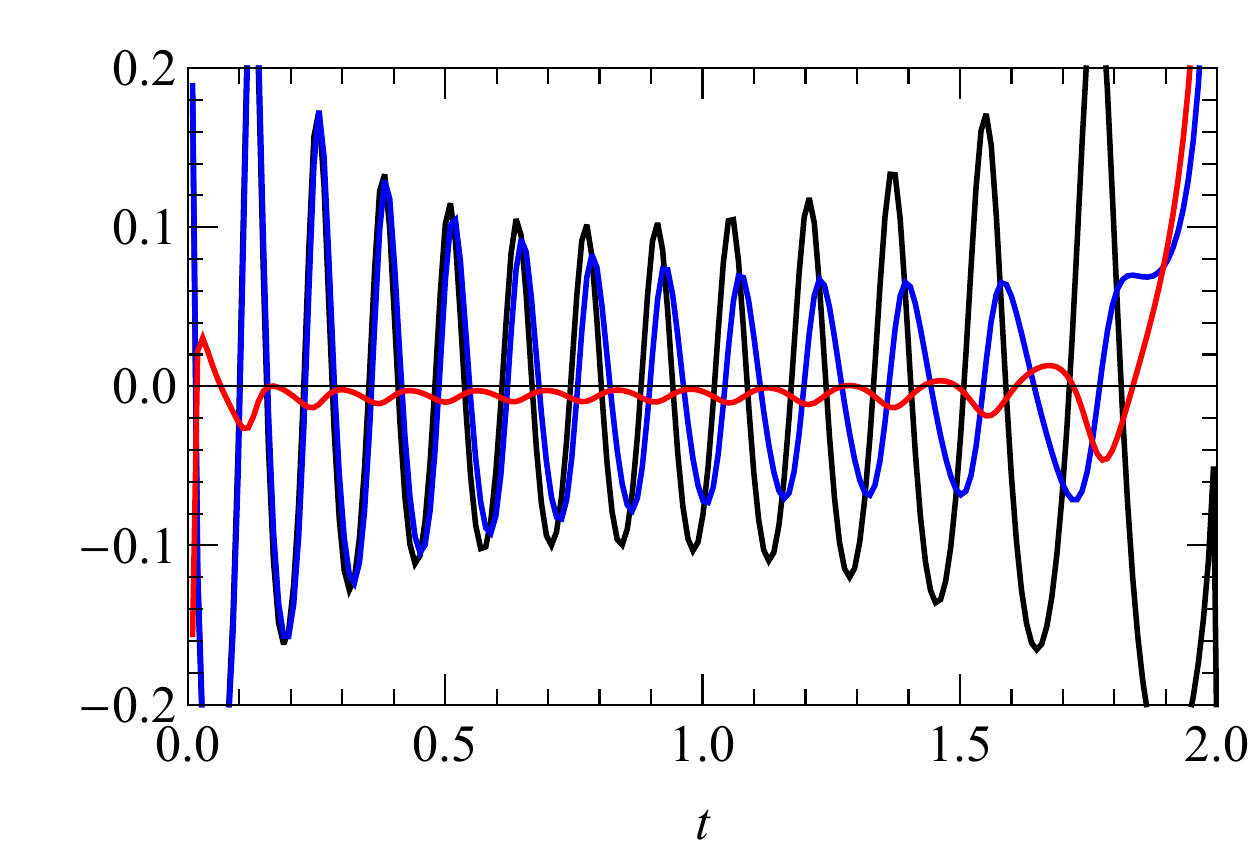}
\includegraphics[width=0.49\textwidth]{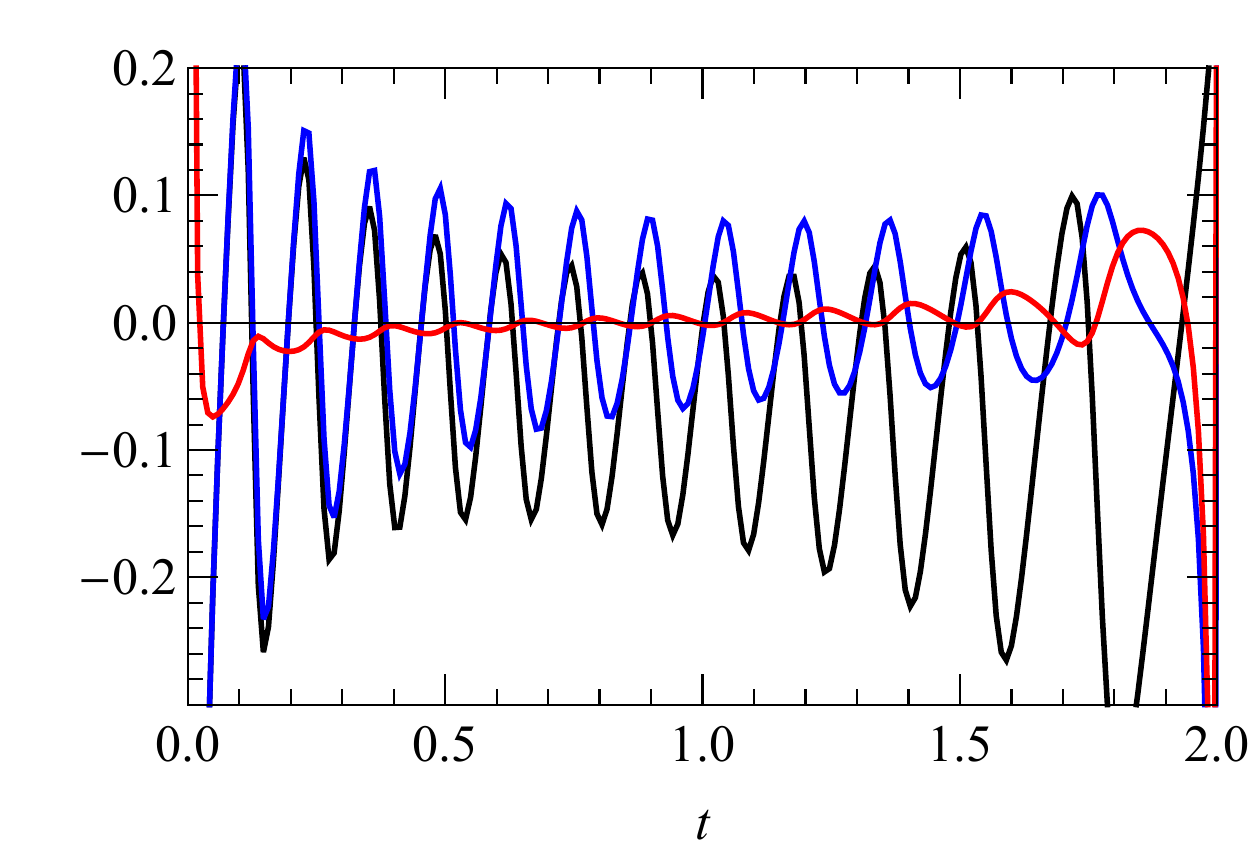}
\end{center}
\caption{Real and imaginary part of $\widetilde{\rho}_{25}^{0.7}(0,t;D) / \rho_{25}^{0.7}(0,t) - 1$ for $D=D_0$ (black), $D=D_1$ (blue) and $D=D_h$ (red).}
\label{fig:k07diff}
\end{figure}

Finally in Fig.~\ref{fig:k07diff}, which refers to $\k=0.7$, we immediately note that both the zero-mode and the $D_1$ truncation are worse than in the previous cases by an order of magnitude. This is explained by the fact that the $(-1,-1)$ term is actually quite close to the first two modes for large $\k$ (see Fig.~\ref{fig:exponents}) and should be consistently included; indeed the truncation $D_h$ provides the same quality of approximation as achieved by $D_l$ or $D_m$ for lower statistical parameter.

\section{Conclusions} \label{sec:concl}

In conclusion, we have presented two approaches to the calculation of the one-body reduced density matrix $\rho_{N}^{\k}(t,t')$ of a gas of $N$ harmonically trapped anyons with repulsive $\delta$-interaction in the Tonks-Girardeau, or hard-core, limit. In the first one we find an exact representation as the determinant of a Hankel matrix of dimension $N-1$; this generalizes the corresponding construction for bosons. In the second approach, we use the replica method with the correct anyonic prescription for the analytic continuation and find a complete asymptotic expansion in analytic form. We showed that, even for relatively small $N$,  the truncation to the first few terms (appropriately chosen according to $\k$) gives an extremely precise approximation (within a few percent except near the $t=t'$ line and when $|t|$ or $|t'|$ get close to the Fermi-Thomas radius). Even though the this result can be in principle improved by performing an ordinary perturbation theory on top of the saddle point integration of Sec.~\ref{sec:replica}, already in the present form it represents a very accurate prediction even for relatively small $N$.

There are several possible generalizations of our work that are worth to be mentioned here.  
First, it could be theoretically interesting and in principle doable to extend our work to a system of hard-core anyons in a finite linear
geometry with Dirichlet or Neumann boundary conditions (the first is physically related to trapping in an infinite square well).
Another promising line of research concerns the study of the non-equilibrium properties of anyon gases. 
There are already a few manuscripts in this direction \cite{campo-08,hc-12,wrdk-14}.
For example, Ref.  \cite{campo-08} generalizes the results for the free expansion of a bosonic Tonks-Girardeau gas 
released from a harmonic trap \cite{mg-05} to the anyonic statistics.
On the same lines, it should be possible to study the behavior of anyonic  observables following the release from a
trap to a finite circle, generalising the bosonic and fermionic results of Ref. \cite{csc-13}.
A more ambitious problem would be to understand the behavior of the anyonic one-body reduced density matrix 
after an interaction quench to the impenetrable limit, as done for the bosonic counterpart in \cite{kcc14}.

\textit{Note added -} 
This paper has taken a very long time to see the light of day.
All the results in Sec.~\ref{sec:hankel} were already present in the master thesis of one of us (MP) 
which dates back to  October 2008, see the link https://etd.adm.unipi.it/t/etd-10012008-115712/. 
When this manuscript was practically completed, the work \cite{Hao-16} appeared, which deals with the same problem as the present work with the approach of Sec.~\ref{sec:hankel}. In fact all the most relevant formulae in that section, Eqs.~(\ref{hankeldet})-(\ref{asymbol2}), have a correspondence in \cite{Hao-16}. Also  Fig.~\ref{fig:xyall} is essentially similar to Figg.1-2 in \cite{Hao-16}, the difference being the choice of the parameters $\kappa,N$.

\section*{Acknowledgments}
GM wants to thank D.~Feder, O.~Lisovyy and S.~Tsuchiya for useful discussions and 
SISSA for hospitality. 
PC acknowledges support from the ERC under the Starting Grant 279391 EDEQS.

\end{document}